\documentclass[draft,grl]{AGUTeX}
 \usepackage[final]{graphicx}
 \usepackage{amssymb}

\usepackage{color}
\newcommand{\unit}[1]{\hat{\vec{#1}}}
\renewcommand{\vec}{\mathbf}

 \authorrunninghead{Fu ET AL.}
 \titlerunninghead{gaps in banded chorus}
 
\begin{document}
\title{Nonlinear sub-cyclotron resonance as a formation mechanism for gaps in banded
chorus}
  
\authors{Xiangrong Fu\altaffilmark{1}, Zehua Guo\altaffilmark{1},
   Chuanfei Dong\altaffilmark{2}, S. Peter Gary\altaffilmark{3}}
  
\altaffiltext{1}{Los Alamos National Laboratory, Los Alamos, NM}
\altaffiltext{2}{Department of Atmospheric, Oceanic, and Space Sciences,
 University of Michigan, Ann Arbor, MI}
\altaffiltext{3}{Space Science Institute, Boulder, CO}

\begin{abstract}
An interesting characteristic of magnetospheric chorus is the presence of a 
frequency gap at $\omega \simeq 0.5\Omega_e$, where $\Omega_e$ is the electron
cyclotron angular frequency. 
Recent chorus observations sometimes show additional gaps near $0.3\Omega_e$ and $0.6\Omega_e$. 
Here we present a novel nonlinear mechanism for the formation of these gaps using
Hamiltonian theory and test-particle simulations in a homogeneous, magnetized,
collisionless plasma. We find that an oblique whistler
wave with frequency at a fraction of the electron cyclotron frequency can
resonate with electrons, leading to effective energy exchange between the wave and
particles.
 \end{abstract} 
 \begin{article}
  \section{Introduction}

Chorus in the Earth's magnetosphere is a type of whistler wave that has been
analysed for decades using ground-based and satellite observations. Intensive
studies have gone into understanding the generation and propagation of these
waves, and how they
affect the magnetospheric plasma \citep[][and references therein]{sazhin_pss_1992}. Chorus is believed to be excited by cyclotron
resonance with anisotropic ($T_\perp > T_\parallel$, where $\perp$ and
$\parallel$ correspond to directions perpendicular and parallel to the
background magnetic field) electrons with energy $>1$ keV injected into the
inner magnetosphere. 
Studies have shown that the resonance between chorus and relativistic
electrons plays an important role in radiation-belt dynamics
\citep[][and references therein]{thorne_grl_2010}. For example, pitch
angle scattering by chorus is a major loss
mechanism for trapped electrons in the outer radiation belt.
Local acceleration due to
interactions between chorus and electrons inside the radiation belt may be a
major mechanism for enhanced relativistic electron population
\citep[e.g.][]{thorne_nature_2013}.

An important feature of chorus is the presence of a gap at one-half the electron
cyclotron frequency in its spectrum, separating two frequency bands (therefore
called ``banded chorus''), a lower band with 0.1 $< \omega /\Omega_e <$ 0.5
and an upper band with 0.5 $< \omega /\Omega_e <$ 0.8
\citep{mered_jgr_2012,li_jgr_2013}, where $\Omega_e$ is the
electron cyclotron angular frequency. Such banded chorus is not
unique to the Earth's magnetosphere, but is also observed in Saturn's
magnetosphere \citep{hospo_jgr_2008}.  Recently, using Cluster spacecraft
measurements,
\citet{macus_ieeex_2014} reported that
sometimes chorus can have more than two bands, with additional gaps near
$0.3\Omega_e$ and $0.6\Omega_e$. Most of these
``multi-banded'' chorus events were observed with oblique wave normal angles
during disturbed geomagnetic conditions. An example of multi-banded chorus
observed by Van Allen Probes A on Feb 10, 2013 is shown in Figure
\ref{fig:obs}.

The narrow gap at $0.5\Omega_e$ between the two bands of chorus has received a
lot of attention since it was discovered (for a review of existing theories, see
\citet{sazhin_pss_1992} and \citet{liu_grl_2011}).  
\citet{liu_grl_2011} hypothesized that two bands of chorus are excited by two
different electron populations with temperature anisotropies through linear
instability, and the hypothesis has been tested in a case study using Van Allen
Probes data by \citet{fu_jgr_2014}.  \citet{schri_jgr_2010} explored the
possibility of generating the lower band through nonlinear wave-wave coupling of
the upper band chorus. \citet{bell_grl_2009} assumed that different bands are
generated in ducts of either enhanced or depleted cold plasma density.
\citet{omura_jgr_2009} explained the gap at $0.5\Omega_e$, where the group
velocity of the whistler wave equals the phase velocity, as nonlinear damping of
a slightly oblique whistler wave packet propagating away from the magnetic
equator, taking into account the spatial inhomogeneity of the magnetic field.
Despite intensive research, a conclusive explanation for banded chorus has not been given yet. 

Existing models for banded chorus have only considered the primary (linear)
resonances when the resonant particles described by unperturbed orbits see a
``time-independent'' wave field. However it was recently found that the nonlinear
resonances which develop by taking into account the perturbed particle motion in the wave
field can also prompt energy exchange between the wave and particles. For
instance, island overlapping due to nonlinear resonances can cause stochastic
ion heating in an oblique Alfv\'en wave with a sub-cyclotron frequency in the
solar corona \citep{chen_pop_2001, guo_pop_2008}. Can the nonlinear resonance
happen between chorus and electrons? If it can, how will it affect the
electrons and chorus? 

Inspired by the interesting feature of nonlinear resonances at sub-cyclotron
frequencies, we propose the following scenario for the banded chorus observation
illustrated in Figure \ref{fig:obs}. A warm (a few hundred eV), anisotropic
($T_\perp>T_\parallel$)
electron velocity distribution drives the whistler anisotropy instability
\citep{gary_pop_2000, gary_pop_2011}, which
gives rise to continuous narrowband ($0.4 \Omega_e<\omega<0.7\Omega_e$)
enhanced magnetic spectra that appear as a relatively coherent temporal waveform. The cold electron (1-100 eV)
response, as will be shown by
both theoretical analysis and test-particle simulations in this work, demonstrates that there is a nonlinear wave-particle interaction whereby
certain electrons come into subharmonic resonance with certain Fourier components. If this
interaction transfers energy from the fluctuations to the electrons, the
resonant Fourier components will be damped and the 
fluctuation spectra will develop gaps at $\Omega_e/2$ and other subharmonics as
shown in Figure \ref{fig:obs}. 

In this paper, we show that, in the absence of primary
resonances, an oblique whistler wave with a frequency at a fraction of
$\Omega_e$ is
able to resonate with the cold electrons nonlinearly, leading to nonlinear
damping/growth of the wave with certain electron distributions. This nonlinear
mechanism, which involves only wave-particle interactions and works in
homogeneous plasmas with a uniform magnetic field, can provide a complementary
element to existing theories on chorus. 
In addition, it can explain additional gaps in chorus spectra
around $0.3\Omega_e$ and $0.6\Omega_e$, as reported recently by
\citet{macus_ieeex_2014}.

The rest of the paper is organized as follows. In Section 2, we present 
a theoretical framework for analysing the dynamics of electrons in an oblique
whistler wave with uniform background magnetic field. The structures of
nonlinear resonances are analysed using Poincar\'e maps by solving the
equations of motion numerically and confirmed by our theoretical calculations
employing the Lie perturbation method.  In Section 3, we show the results of
test-particle simulations for an ensemble of electrons with certain velocity
distributions. The effects of nonlinear resonances on the electron distribution
function and the total kinetic energy are investigated.
In Section 4, we
discuss how this nonlinear mechanism is related to frequency gaps in
magnetospheric chorus. Finally, conclusions are given in Section 5.

 \section{Theoretical Analysis}
In this section, we analyse the dynamics of electrons in a single oblique
whistler wave and an uniform background magnetic field using Hamiltonian theory.
 \subsection{Hamiltonian}
For simplicity, we consider a uniform plasma in a uniform background magnetic
field, $\vec{B}_0 = B_0 \unit{z}$, and the whistler wave dispersion relation in
the cold plasma limit is given as \citep[][Equation (2-45)]{stix_book_1992} 
\begin{equation}
\left(\frac{ck}{\omega}\right)^2 = 1 - \frac{\omega_{e}^2}{\omega(\omega - \Omega_e\cos\theta)}\,,
\end{equation}
where $\omega_{e}$ is the plasma
frequency, $k$ is the wave number, and the
whistler wave is oblique with a small wave normal angle $\theta$ with respect to
the background magnetic field so that
$\alpha\equiv \tan \theta = k_x/k_z$. Assuming $\theta \ll 1$ and using the cold
plasma theory \citep{stix_book_1992}, the vector potential can be written as
\begin{equation}
\vec{A} = B_0 x \hat{\vec{y}} + \epsilon (B_0/k_z) (\sin\Psi\unit{x} + \cos\theta \cos\Psi\unit{y}) \,,
\end{equation}
where $\Psi = k_x x + k_z z - \omega t$ is the phase of the wave and
$\epsilon\equiv B_w/B_0$ denotes the perturbation magnitude. To eliminate
electric fields, we move to the wave frame by a transformation, $\vec{x}' =
\vec{x} - (\omega/k_z) t\unit{z}$. Normalizing time to $1/\Omega_e$, 
magnetic field to $B_0$, mass to $m_e$ and length to $1/k_z$, the normalized Hamiltonian for
the electron is given by
\begin{equation}
H =  \frac{1}{2}\left[p_x^2 + (p_y + x)^2 + p_z^2\right] + \epsilon \left[
p_x\sin\psi + (p_y + x) \cos\theta\cos\psi\right] 
 + \frac{1}{2}\epsilon^2 (1 -  \sin^2\theta\cos^2\psi)\,,
\label{eq:H}
\end{equation}
where $p_x$, $p_y$, and $p_z$ are the canonical momenta and $\psi = \alpha x + z$. In
the wave frame, the electron energy is conserved as $H$ does not depend on time
explicitly and $p_y$ is a constant of motion since $H$ is independent of $y$.
A set of equations of motion for the electron can be readily obtained from the
Hamilton's equation \citep[e.g.][]{jose_book_1998}: 
 \begin{eqnarray}
\label{eq:eom1}
   \dot{x}&=&p_x+\epsilon \sin\Psi,\\
   \dot{p}_x&=&-(p_y+x)-\epsilon[\alpha p_x\cos\Psi-\alpha(p_y+x)\cos\theta
 \sin\Psi+\cos\theta\cos\Psi]\nonumber\\
 &&-\frac{\epsilon^2}{2}\alpha \sin^2\theta\sin(2\Psi),\\
 \dot{z}&=&p_z,\\
 \dot{p}_x&=&-\epsilon[p_x\cos\Psi-(p_y+x)\cos\theta\sin\Psi]-\frac{\epsilon^2}{2}\sin^2\theta\sin(2\Psi),
\label{eq:eom4}
 \end{eqnarray}
where dots represent time derivatives.

Without loss of generality, we choose the following parameters
 relevant to chorus in the Earth's magnetosphere.
The whistler wave satisfies
 $\omega=0.4\Omega_e$,
 $k=0.9\omega_{e}/c$ and
 $\theta=26.6^\circ$ ($\alpha=\tan \theta=0.5$), where $c$ is the speed of
light and $\omega_{e}/\Omega_e=5$. So the parallel phase speed
 of the wave is $\omega/k_z\approx 0.1c$, and the energy is 
 normalized to $m_e(\Omega_e/k_z)^2 \approx 0.06 m_ec^2=30.66$ keV.
 Note that the choice of the whistler wave frequency is for illustration
purpose. It will be shown later that, for waves closer to $0.5\Omega_e$, the
difference is only in the energy of resonant electrons. 
The extension to a whistler wave packet with multiple Fourier components is also
straightforward \citep{lu_apj_2009}.
Unless otherwise specified, the above parameters are used in the rest of the paper. 

 \subsection{Poincar\'e maps}
\begin{figure}
   \centering
   \includegraphics[width=0.6\textwidth]{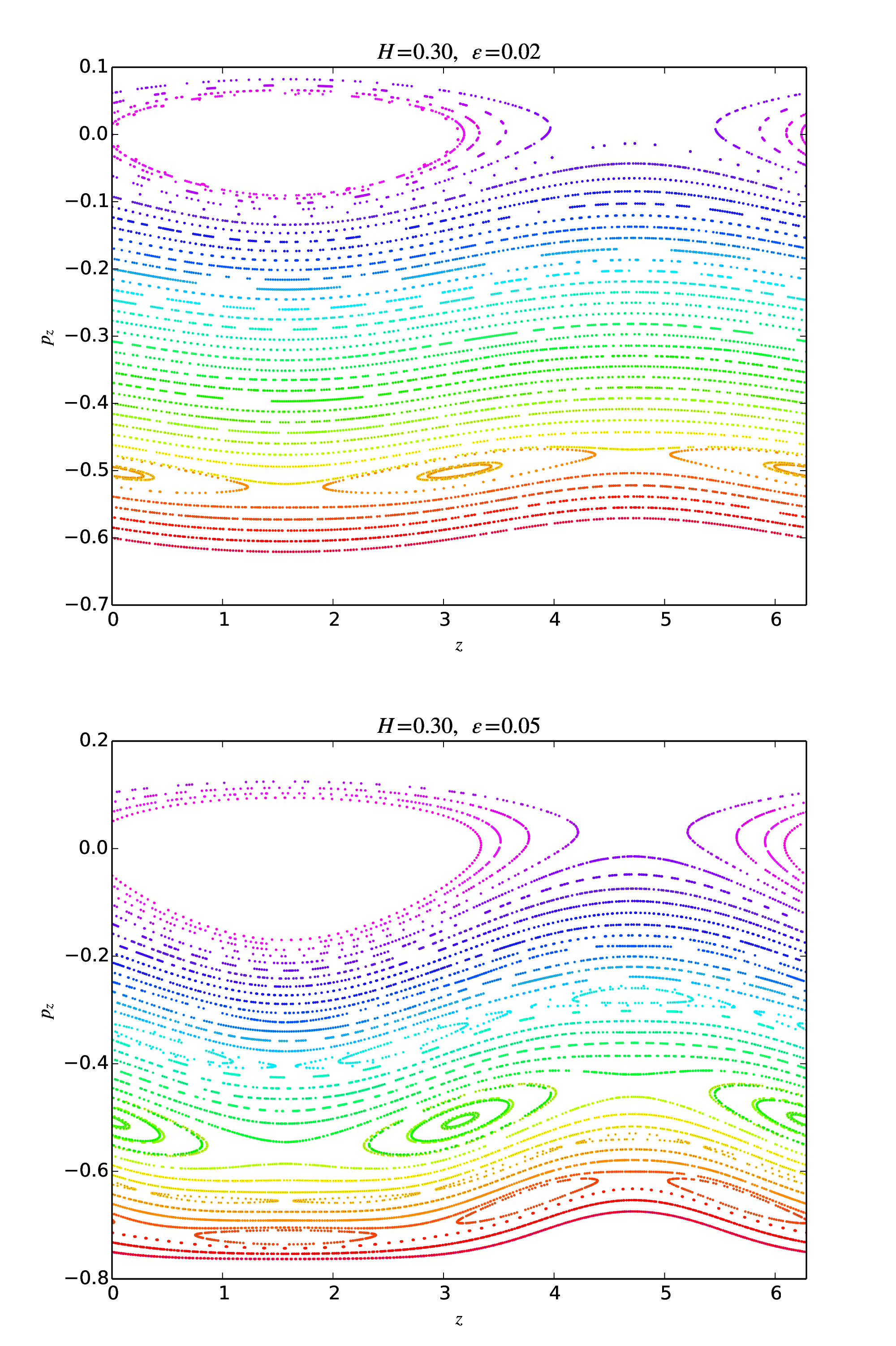}
   \caption{(top) Poincar\'e map in $(p_z,z)$ plane for electrons in the presence of
an oblique whistler wave with an amplitude $B_w/B_0=0.02$. The energy of
electrons in the wave frame is fixed with $H=0.3$. Islands develop near
$p_z=0$ and $p_z=-\frac{1}{2}$. (bottom) Similar Poincar\'e
map, but with a larger wave amplitude $B_w/B_0=0.05$. Additional
resonant islands develop near $p_z= -\frac{1}{3}$ and $p_z=-\frac{2}{3}$.}
   \label{fig:poinc}
 \end{figure}

 In our Hamiltonian model, electrons are moving in a 4-dimensional phase space
 $(x, p_x, z, p_z)$. 
 With Poincar\'e surfaces of section (or maps), one can visualize the
 wave-particle resonances in phase space. We construct a Poincar\'e map in $(p_z, z)$ by
 recording points when the trajectory of an electron in phase space crosses the surface of $x=0$ with $p_x>0$.
 Electrons are initialized with $x=0$, $p_y=0$, fixed $H$ and a range of parallel velocities $p_z$'s.
 In the top panel of Figure \ref{fig:poinc}, a map for electrons with energy $H=0.3$
 is shown in the presence of an oblique whistler wave with $\epsilon=0.02$. 
 In the map, a major island in $(p_z,z)$ plane located at $p_z=0$ demonstrate
trapping of particles 
 by the wave satisfying $\omega-k_zv_z=0$ (note $p_z=m_e(v_z-\omega/k_z)$ before
normalization), which
 is the well-known Landau resonance
 condition. Another set of two islands develop at $p_z=-\frac{1}{2}$, which
 corresponds to a new resonance condition 
\begin{equation}
\omega-k_zv_z=\Omega_e/2,
\label{eq:nl_condition}
\end{equation} 
suggesting that
 electrons can resonate with finite amplitude oblique whistler waves at
half cyclotron frequency.
 A similar mechanism has been proposed for the sub-cyclotron resonance
 between ions and oblique Alfv\'en waves \citep{chen_pop_2001, guo_pop_2008}. 
 With higher energy, the primary cyclotron resonance
 between electrons and the whistler wave can be seen from the island located at
 $p_z=-1$ (not shown).
\subsection{Sub-cyclotron resonances}
We can show that the half cyclotron resonance is the result of nonlinear
dynamics by taking into account the perturbed electron orbit in the presence of
wave fields. The analysis is greatly simplified in the so-called guiding-center
coordinates. Through a canonical transformation \citep{guo_pop_2008}, the Hamiltonian becomes 
\begin{equation}
H = H_0 + \epsilon H_1 + \epsilon^2 H_2
\end{equation} 
where $H_0 = J + p_z^2/2$ is the unperturbed guiding-center Hamiltonian and
\begin{eqnarray}
H_1 &=&  \sqrt{2J}\left[ \cos\phi \sin (\alpha \sqrt{2J}\sin\phi + z) + \sin\phi \cos\theta\cos(\alpha \sqrt{2J}\sin\phi + z)\right]\,,\\
H_2 &=& - \frac{1}{2}\sin^2\theta\cos^2(\alpha \sqrt{2J}\sin\phi + z)\,.
\end{eqnarray}
Here, $J=[(p_y+x)^2 + p_x^2]/2$ denotes the perpendicular energy and $\phi$ the
gyro-angle. Interestingly, if the whistler wave is purely parallel to the
background magnetic field ($\theta = 0$), $H_2$ vanishes and $H_1 = \sqrt{2J}
\sin(z - \phi)$ contains only the primary resonance. So the electron dynamics
becomes integrable and there will not be any nonlinear resonance. Due to the dependence of the perturbed Hamiltonian on $\phi$ and $z$, both $J$ and $p_z$ vary in time. These variations are related to the magnetic drift of the guiding center which can then give rise to nonlinear resonances via the $\vec{k_\perp} \cdot \vec{X}_g$ term with $\vec{X}_g$ being the magnetic drift. 

To analyse the perturbed Hamiltonian, we introduce the Lie transform
method \citep[e.g.][]{jose_book_1998} which utilizes particular canonical transformations to obtain
perturbation series in terms of Poisson brackets. Similar to the approach in
\citet{guo_pop_2008}, a generating function
\begin{equation}
W_1(\phi, z, J, p_z) =  \sqrt{2J}\sum_n (\sin^2\frac{\theta}{2} J_{n+1} + \cos^2\frac{\theta}{2} J_{n-1}) \frac{\cos( n \phi + z)}{n + p_z}\,,
\label{eq:w1}
\end{equation}
is obtained to remove the first order perturbations by setting $[W_1, H_0] + H_1
= 0$ with $[\,,\, ]$ being the Poisson bracket. This generating function
indicates singularities when $p_z = -n$, with $n$ being any integer. In the
laboratory frame, they simply correspond to the well known cyclotron resonance
condition $ \omega - k_{z} v_{z} - n \Omega_e = 0 $ when $n\neq 0$ and the
Landau resonance condition $\omega -k_z v_z = 0$ when $n=0$. 

After the Lie transformation, the new Hamiltonian becomes
$
H^{\prime} = H_0 + \epsilon^2 (H_2 + [W_1, H_1]/2) + O(\epsilon^3)
$,
where the second order perturbation reads
\begin{equation}
H_2^\prime = \frac{1}{4}\left[\sum_{m,n} ( A_{m,n} - \sin^2\theta J_mJ_n)
\cos(l\phi + 2z) - \sum_{m, n} ( B_{m,n} + \sin^2\theta J_mJ_n)
\cos(m-n)\phi\right]\,,
\label{eq:h2} 
\end{equation}
with $l = m+n$. The functions $A_{m,n}(J, p_z), B_{m,n}(J,p_z)$ are introduced
for simplicity. Their explicit expressions are tediously long and will be
presented later in a separate paper.  
The new Hamiltonian clearly shows the existence of nonlinear resonances when the
phase $(l\phi + 2z)$ remains a constant or $p_z = - l/2$. In the laboratory
frame, they correspond to the second order resonance conditions $\omega - k_zv_z
-(l/2)\Omega_e = 0$. The second term in Equation (\ref{eq:h2}) does not contribute to
resonances. Therefore, our analytical calculation confirms the existence of the
nonlinear resonance at the half cyclotron frequency ($l=1$). Following the
result in \citet{guo_pop_2008}, we know that the island width of second order
resonances is proportional to $\epsilon$ while it is $\sqrt{\epsilon}$ for the
primary resonances. This means the damping due to second order resonance is
weaker than the primary resonance by square root of the wave magnitude.
Continuing to the third order expansion, we will obtain resonances at $p_z =
-l/3$ which is present in the bottom panel of Figure \ref{fig:poinc} where
the wave has a larger wave amplitude.

\section{Test particle simulations}

A wave-particle resonance causes efficient energy exchange
between the wave and resonant particles. For example, a wave is damped via
Landau resonance while the resonant particles gain same amount of kinetic energy
the wave loses. To further illustrate how the nonlinear resonance, at $\omega -
k_z v_z - \Omega_e/2 = 0$, can damp waves around the half cyclotron frequency,
we perform test-particle simulations of a large number of electrons and investigate the changes in the electron distribution and their
kinetic energy in the presence of an oblique whistler wave.

\begin{figure}
   \centering
   \includegraphics[width=0.6\textwidth]{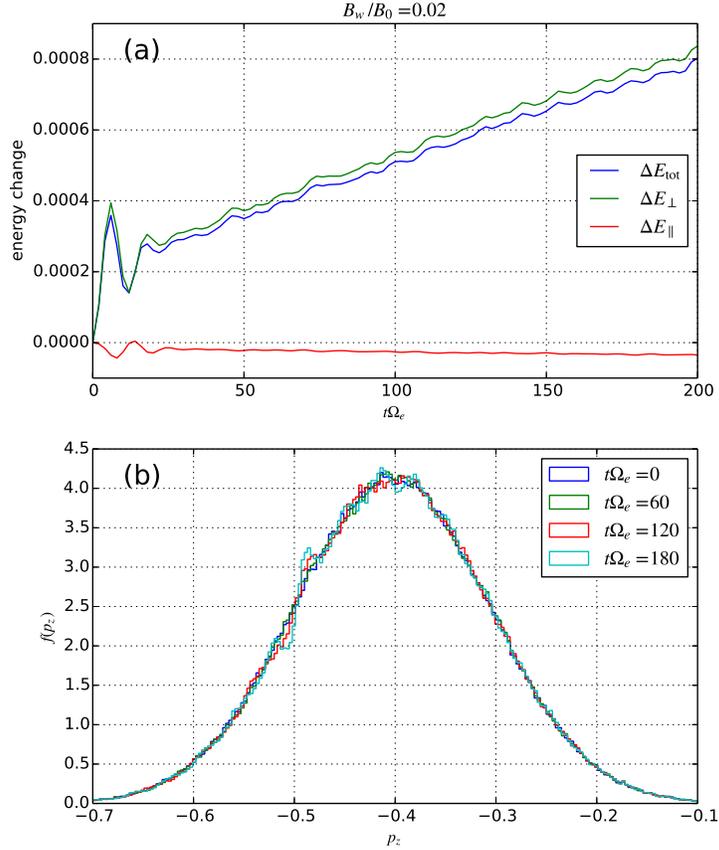}
   \caption{(a) Change of the averaged energies per test electron in the presence of an oblique
whistler wave with an amplitude $B_w/B_0=0.02$. The averaged initial total energy per
electron $E_{\rm tot,0}=0.028$. (b) The distributions of $f(p_z)$ at different times of the
simulation. Changes of $f(p_x)$ and $f(p_y)$ are minor.
}
   \label{fig:energy}
 \end{figure}

In our test-particle simulation, about $10^{6}$ electrons with Maxwellian velocity distribution in
the lab frame, 
\begin{equation}
f(v_x,v_y,v_z)=f_0\exp\left[-m_e\left(\frac{v_x^2+v_y^2}{2T_\perp}+\frac{v_z^2}{2T_\|}\right)\right],
\end{equation}
are loaded initially.
Here we choose electron temperatures $T_\perp=T_\parallel \sim 300$ eV so that
$v_t=\sqrt{T_\|/m_e}\approx|(\omega-\Omega_e/2)/k_\||$.
Electrons are advanced according to equations of motion given by Equations.(\ref{eq:eom1})-(\ref{eq:eom4}).
In the presence of an oblique whistler wave with wave amplitude $B_w/B_0=0.02$, 
the energy (calculated in the lab frame) evolution of these electrons is shown in 
Figure \ref{fig:energy}a.
After initial oscillations, the averaged energy per electron ($E$) increases at a rate
$\Delta E/(E_0\Delta t)\approx 10^{-4}\Omega_e$ (where $E_0$ is the averaged initial
energy per electron) and the energy gain is predominantly in the perpendicular
direction. In a self-consistent simulation where the wave field
follows Maxwell equations and the total energy of the system is conserved, the
increase of the electron energy must come from the decrease of wave energy. This
implies that the wave will be damped by the nonlinearly resonant electrons. The
rate of particle energy gain normalized to the wave energy is $2\mu_0 n_em_e
(\Omega_e/k_z)^2\Delta E /(B_w^2\Delta t)\approx 0.02\Omega_e$.

The electron velocity distributions in the wave frame at $t\Omega_e=0,60,120,180$ are shown in
Figure \ref{fig:energy}b. Clearly, the parallel velocity distribution $f(p_z)$
deviates from the initial Maxwellian in the vicinity of $p_z=-0.5$, which
corresponds to the second order resonance shown in 
the Poincar\'e map in Figure \ref{fig:poinc}. The changes in both $f(p_x)$ and
$f(p_y)$ are small (not shown).
For the given Maxwellian distribution function, the nonlinear cyclotron
damping steepens the gradient of $f(p_z)$ around $p_z=-0.5$, causing the parallel kinetic energy to
decrease, and the perpendicular kinetic energy to increase (see Figure
\ref{fig:energy}b and
\ref{fig:energy}c). This is similar to the primary cyclotron resonance where
the electrons are scattered along the constant energy surfaces in the wave frame
as pointed out by \citet{kennel_pof_1966}. 

Just like the Landau or cyclotron resonance, the nonlinear resonance can lead
to damping/growth of the waves at sub-cyclotron frequencies when the electron
distribution satisfies certain conditions (the stability criteria). Even though
the particle motion is described as periodic in our theoretical analysis and
test-particle simulations, what determines the wave damping/growth rate is how
fast the resonant particles can take (give) energy from (to) the perturbed fields.
This is the same as in Landau or cyclotron damping, which requires the resonant
particles to gain energy at a rate faster than their bounce frequency in the
perturbed field. The wave begins to saturate when this condition is violated.
The difference in the nonlinear damping mechanism is that the relevant bounce
motion is now described by the second order perturbed Hamiltonian and thus
becomes much slower than the linear bounce motion. A test-particle simulation
describes the particle dynamics at the beginning stage of the wave-particle
interaction and can predict the rate of energy gain or loss by resonant
particles depending on the distributions, as discussed in the following.

\begin{figure}
   \centering
   \includegraphics[width=0.6\textwidth]{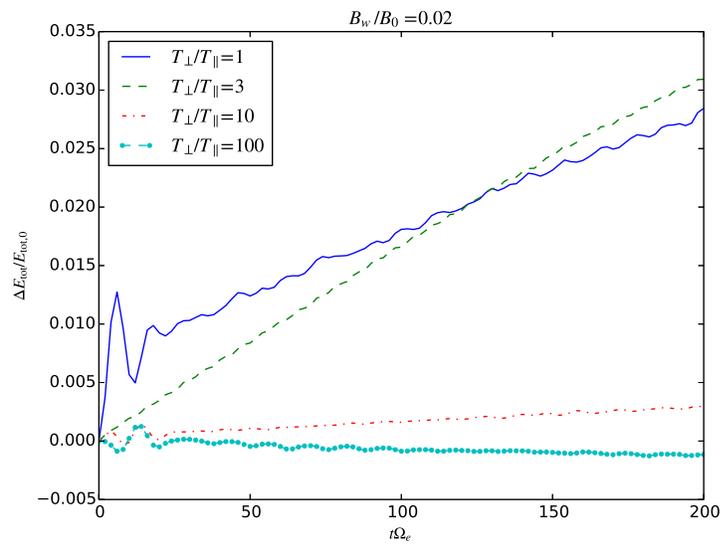}
   \caption{Relative change of energies per electron for cases with temperature
anisotropy $T_\perp/T_\parallel=1,3,10,100$. All parameters except
$T_\perp$ are fixed.}
   \label{fig:anisotropy}
 \end{figure}

When the temperature anisotropy $T_\perp/T_\parallel$ is strong
enough, the nonlinear resonance may also lead to decrease of the particle energy
and thus instability of the whistler wave.
This is
demonstrated by a comparison of runs with different initial anisotropies, as shown in Figure
\ref{fig:anisotropy}. Same parameters as those in Figure \ref{fig:energy} except
the perpendicular
temperature are used in these runs. For $T_\perp/T_\parallel=100$, the electrons lose energy,
which implies wave growth in a self-consistent situation. Interestingly, for
anisotropies well below the instability threshold, the temperature anisotropy
can enhance the nonlinear wave-particle resonance, as shown by the case with
$T_\perp/T_\parallel=3$ initially. This is consistent with our analysis as in
Equation (\ref{eq:w1})
where a factor of $\sqrt{2J}$ shows the dependence on the particle perpendicular energy. 

\begin{figure}
   \centering
   \includegraphics[width=0.6\textwidth]{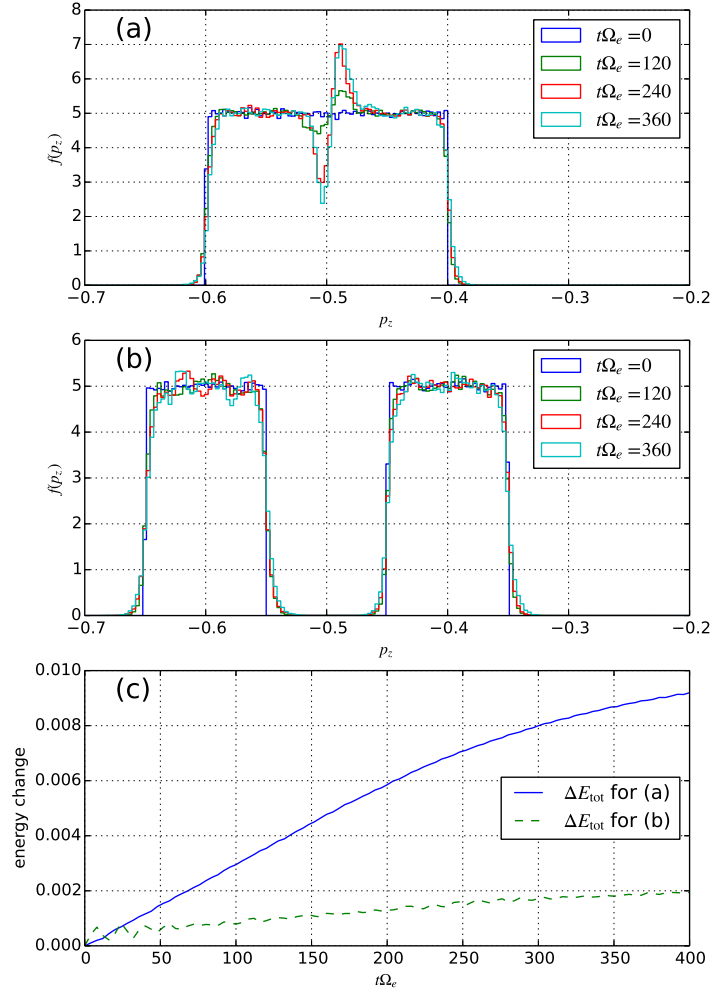}
   \caption{The evolution of the distribution function $f(p_z)$ for (a) an
initial flat-top
distribution, and (b) a flat-top distribution with resonant particles removed. The
comparison of the change of total energy per electron is shown in (c).}
   \label{fig:flat}
 \end{figure}

To further illustrate the sub-cyclotron resonance and clarify that the energy gain of electrons is not dominated by
numerical heating or the pseudo-heating \citep{dong_pop_2013} in our simulations, we carry out two comparison test cases. In
the first case, we artificially load electrons with  $f(p_z)$ flat
in the range $-0.6<p_z<-0.4$ as shown in Figure \ref{fig:flat}(a). Due to the
nonlinear resonance near half cyclotron frequency, electrons gain energy from the wave. In the second case,
we remove the resonant particles within $-0.55<p_z<-0.45$ from the distribution as shown in Figure
\ref{fig:flat}(b). The resulting energy gain in the lab frame (Figure \ref{fig:flat}c) is significantly reduced. Therefore it is clear that the dominant energy gain of electrons is caused by the nonlinear resonance.

\section{Application to Banded Chorus}
In the Earth's magnetosphere, cold plasmaspheric electrons ($\sim$ 1 eV)  extend
into a region called the
``plasma trough'' with $L=4-8$ ($L$ is the equatorial distance of a magnetic
field line from the center of the Earth
in the unit of Earth's radius) \citep[e.g.][]{carpen_jgr_1992}, where they meet
$\sim$ 1 keV electrons from the plasma sheet. Banded
chorus is frequently observed in this region \citep[e.g.][]{mered_jgr_2012}.  We conjecture
that the gap of the banded chorus may be caused by the nonlinear damping of oblique
whistler waves by these cold electrons with energy 1-100 eV.
For 10 eV electrons, whose thermal velocity $v_t\sim 6\times 10^{-3}c$,
the nonlinear resonance condition (Equation \ref{eq:nl_condition}) becomes
$\omega \approx 0.5\Omega_e$, i.e., these electrons could damp oblique
whistler waves with frequency near $0.5\Omega_e$, leaving a gap in the wave
power spectrum. 

Wave analysis often
shows chorus is quasi-parallel, but we have also shown that even
with a wave normal angle of $26.6^\circ$, nonlinear wave-particle interactions can be
significant. Furthermore, even if the chorus is excited purely parallel in the 
source region close to the magnetic equator, as the wave propagates away from
the equator, it will soon have an oblique component due to the curved nature of
Earth's dipole-like magnetic field. 

Here, we have presented cases with a modest wave amplitude $B_w/B_0 = 0.02$
to better illustrate the nonlinear resonance mechanism with less computational
constrains. The numerical integration scheme can introduce errors into the
particle's energy, therefore, for smaller wave amplitudes, we need to both
reduce the time step to suppress the numerical heating and increase number of
particles to resolve the resonance structure in phase space. Typical amplitude of
chorus in the magnetosphere is about $B_w/B_0\sim 0.001$, but large
amplitude whistler waves have also been reported with $B_w/B_0>0.01$
\citep[e.g.][]{santo_grl_2014}. The sub-cyclotron resonance can be strong and
effective in damping these large amplitude whistler waves as shown in the
previous section. For smaller wave amplitude, we have also carried out
test-particle simulations with $B_w/B_0=0.002$, $\omega=0.46\Omega_e$,
$T_\parallel=50 eV$, and $T_\perp=100 eV$. The results (not shown here) indicate
that the same nonlinear resonance occurs, although the rate of the electron
energy gain is reduced ($\Delta E/(E\Delta t)\approx 6\times 10^{-6}\Omega_e$) due to
the lower wave amplitude. Therefore the proposed nonlinear damping mechanism at
sub-cyclotron frequencies is robust and can play a role for waves observed in
the magnetosphere. Furthermore, in the presence of multiple
whistler modes (or a narrow band) near half cyclotron frequency, the nonlinear damping is expected
to be enhanced due to the overlapping of resonances as suggested by
\citet{chen_pop_2001} and \citet{lu_apj_2009}. This effect will be addressed in a separate paper.

If the wave amplitude increases to $B_w/B_0=0.05$, additional resonant islands
near $p_z=-1/3$ and $p_z=-2/3$ develop in the Poincar\'e map (Figure \ref{fig:poinc}, bottom panel), which may explain
the additional gaps near $0.3\Omega_e$ and $0.6\Omega_e$ observed by
\citet{macus_ieeex_2014}. As \citet{macus_ieeex_2014} reported, multi-banded
chorus were observed during more disturbed times with the average Kp $\sim 3$,
larger than the average Kp $\sim 2$ for chorus observed with one or two bands.
These higher order nonlinear resonances, as well as
the half cyclotron resonance we have shown, belong to a set of nonlinear resonant conditions below
the cyclotron frequency, $\omega-k_\|v_\|=N\Omega_e/M$, where $M,N$ are 
integer and $N<M$ \citep{guo_pop_2008}.

 \section{Conclusions}
\begin{figure}
   \centering
   \includegraphics[width=0.8\textwidth]{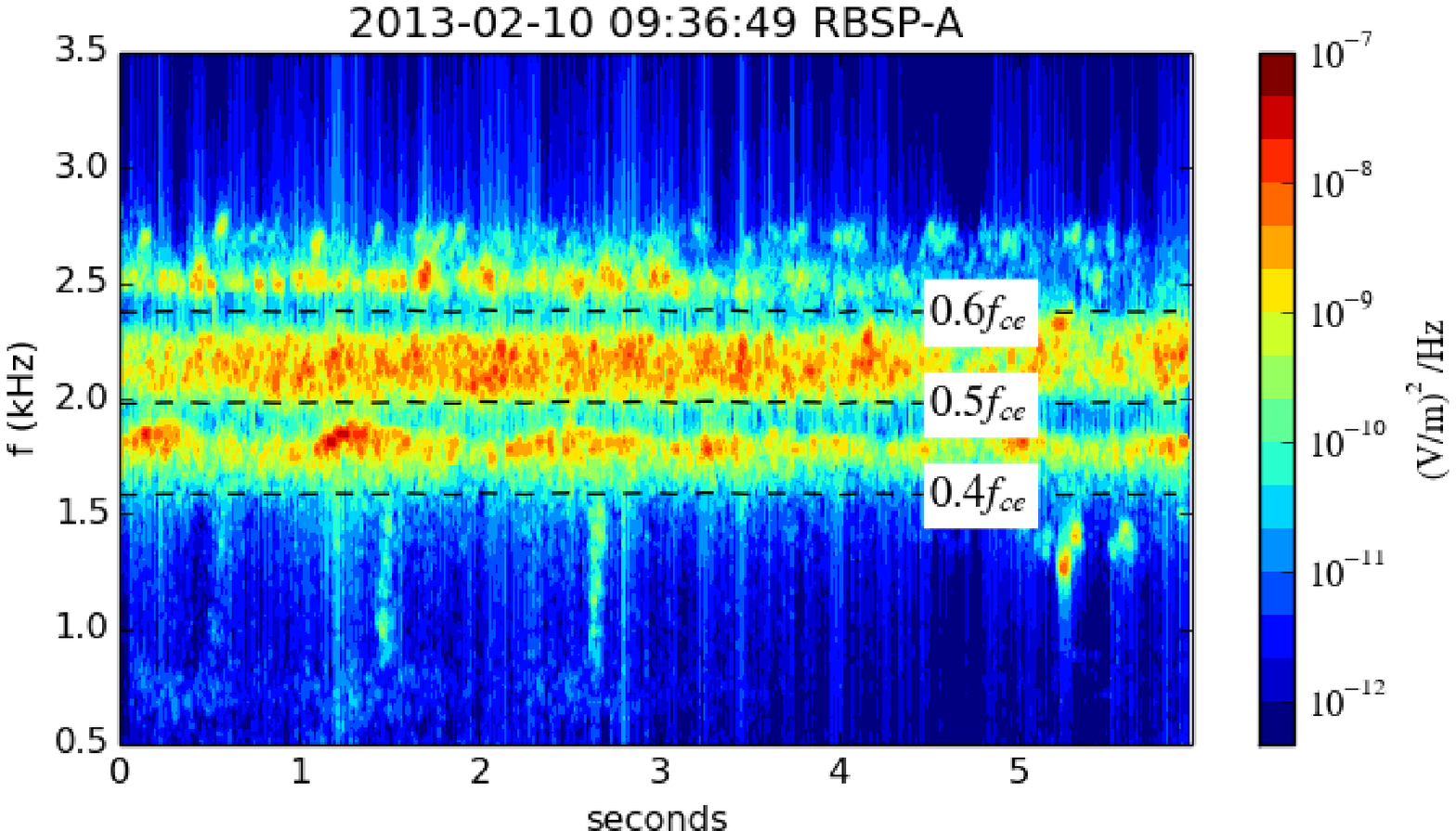}
   \caption{A multi-banded chorus measured by Van Allen Probes A on Feb 10,
     2013: the magnetic field spectrogram with two frequency gaps at 0.5$f_{ce}$ and 0.6$f_{ce}$, 
where $f_{ce}$ is the local electron gyrofrequency. Black dashed lines indicate 0.4$f_{ce}$,
   0.5$f_{ce}$ and 0.6$f_{ce}$, respectively. }
   \label{fig:obs}
 \end{figure}

In this paper, we have presented a nonlinear resonant mechanism between an oblique
whistler wave and electrons, satisfying the resonant condition
$\omega-k_\|v_\|=\Omega_e/2$, by theoretical analysis and test-particle
simulations. Our mechanism works in a homogeneous plasma with a constant
background magnetic field, and may explain the frequency gap at $0.5 \Omega_e$
frequently observed in the
power spectra of magnetospheric chorus. Further more, similar nonlinear
resonances may explain the frequency gaps at $0.3 \Omega_e$ and $0.6\Omega_e$ in
chorus observations recently reported by \citet{macus_ieeex_2014} and as shown
in Figure \ref{fig:obs}.
This mechanism provides a complementary element to existing theories on banded chorus.
The detailed theoretical analysis (Section 2) will be presented in a separate paper
later. The ability of this nonlinear mechanism to explain frequency gaps in
chorus emissions needs to be further investigated in a self-consistent way (e.g.
via particle-in-cell simulations) to address the dependence of damping rates on
various plasma and wave parameters.
  
 \end{article}
\acknowledgement{The authors acknowledge William Kurth, George Hospodarsky,
Craig Kletzing and
the Van Allen Probes EMFISIS team for providing wave data. The data are 
available on the EMFISIS website ({\tt https://emfisis.physics.uiowa.edu/data/index}).
This research was supported in part by the
National Aeronautics and Space Administration and by the Laboratory
Directed Research and Development program of Los Alamos National Laboratory.
LA-UR-15-20444.
}

 \end{document}